\documentclass[prb, aps, twocolumn, floatfix, superscriptaddress]{revtex4-2}
\usepackage{amsmath}
\usepackage{amssymb}
\usepackage{graphicx}
\usepackage{color}

\newcommand{\be}{\begin{equation}}
\newcommand{\ee}{\end{equation}}
\newcommand{\emb}{${\rm EuMg_2Bi_2}$}
\newcommand{\ems}{${\rm EuMg_2Sb_2}$}

\newcommand{\esa}{${\rm EuSn_2As_2}$}
\newcommand{\eg}{${\rm EuGa_4}$}
\newcommand{\ea}{${\rm EuAl_4}$}
\newcommand{\bea}{\begin{eqnarray}}
\newcommand{\eea}{\end{eqnarray}}
\newcommand{\bse}{\begin{subequations}}
\newcommand{\ese}{\end{subequations}}

\begin{document}

\title{Temperature-dependent Eu spin reorientations in the tetragonal A-type antiferromagnet EuGa$_{\bf 4}$ induced by small {\it ab}-plane magnetic fields}

\author{Santanu Pakhira}
\affiliation{Ames National Laboratory, Iowa State University, Ames, Iowa 50011, USA}
\affiliation{Present Address: Institute for Quantum Materials and Technologies, Karlsruhe Institute of Technology, D-76021 Karlsruhe, Germany}
\author{David C. Johnston}
\affiliation{Ames National Laboratory, Iowa State University, Ames, Iowa 50011, USA}
\affiliation{Department of Physics and Astronomy, Iowa State University, Ames, Iowa 50011, USA}

\date{\today}

\begin{abstract}

The body-centered-tetragonal antiferromagnet \eg\ exhibits A-type antiferromagnetic order below its N\'eel temperature $T_{\rm N} = 16.4$~K in magnetic field $H = 0$ where the moments are ferromagnetically aligned in the $ab$-plane with the Eu moments in adjacent Eu planes aligned antiferromagnetically. Previous magnetization versus field $M_{ab}(H)$ measurements revealed that the moments exhibit a spin-reorientation transition at a critical field $H_{c1}$ where the Eu moments become perpendicular to an in-plane magnetic field  while still remaining in the $ab$~plane.  A theory for $T=0$~K was  presented that successfully explained the observed low-field moment-reorientation behavior at $T = 2$~K\@.  Here we present a theory explaining the observed $T$ dependence of $M_{ab}(H,T<T_{\rm N})$ in the [1,0,0] direction for $H\leq H_{\rm c1}(T)$ from 2 to 14~K arising from a $T$-dependent anisotropy energy.

\end{abstract}

\maketitle

\begin{figure}
\includegraphics [width=2.8in]{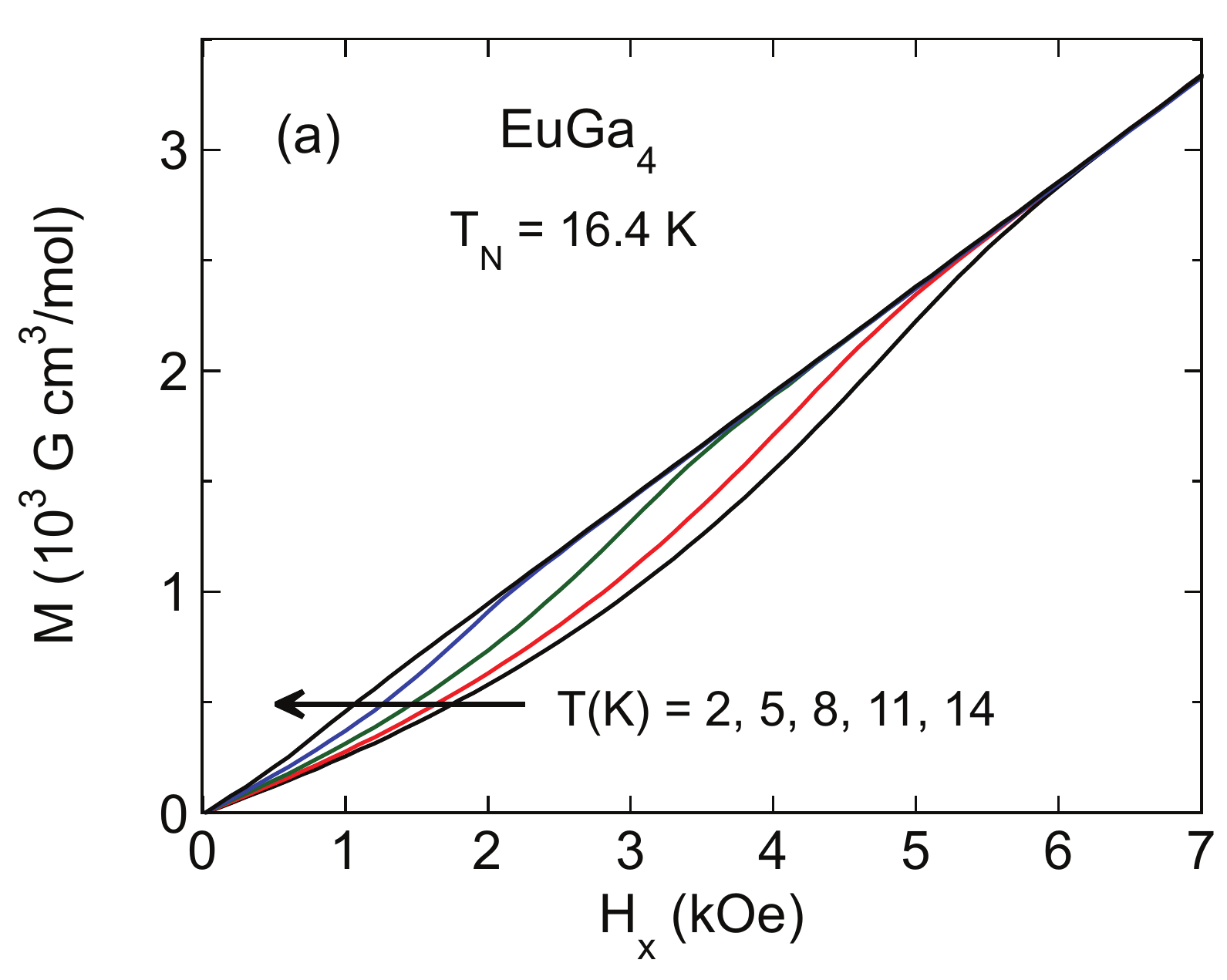}
\includegraphics [width=2.8in]{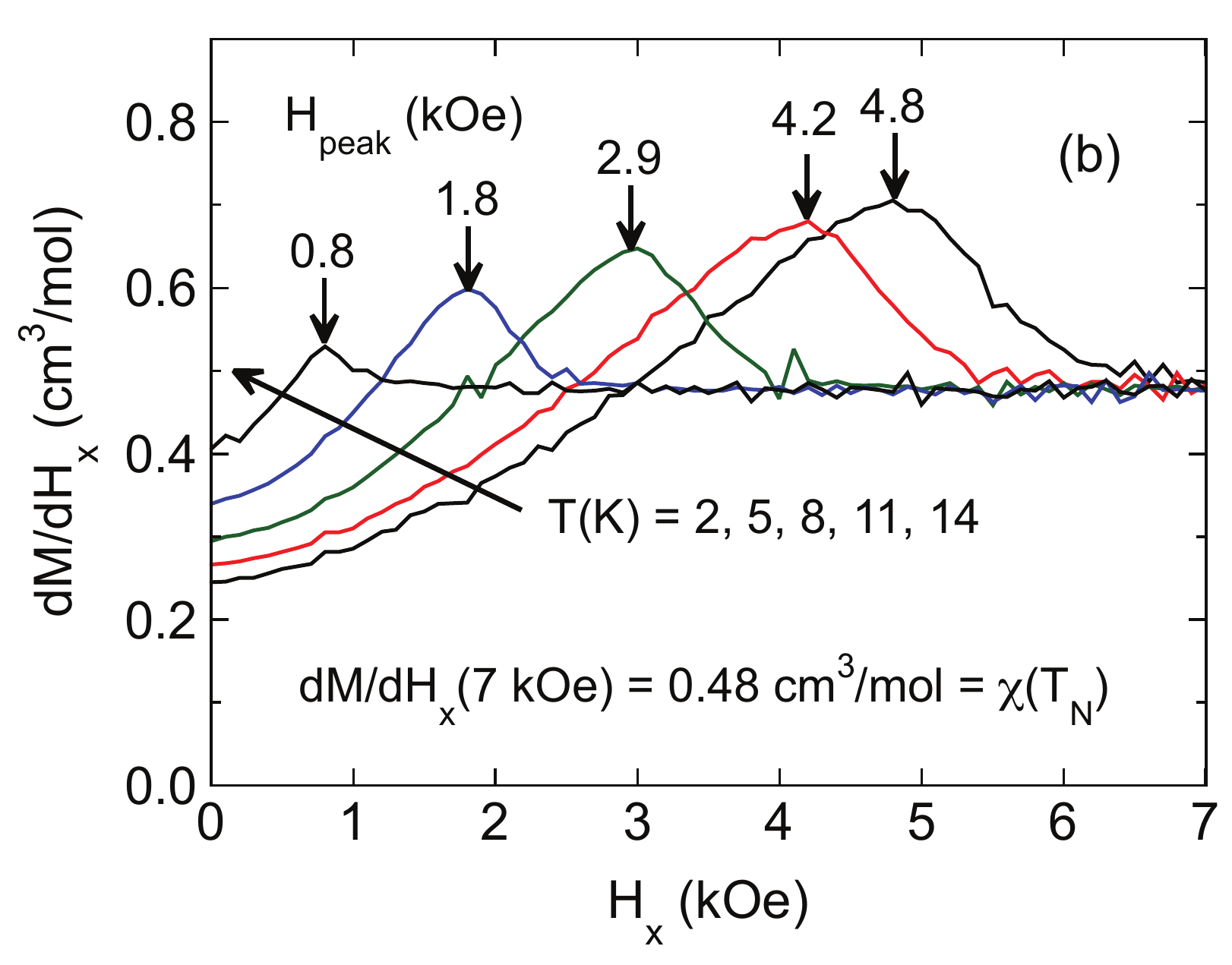}
\caption {(a)~Magnetization~$M$ versus $ab$-plane magnetic field $H_x \parallel\,[1,0,0]$ for \eg\ with $T<T_{\rm N} = 16.4$~K~\cite{Pakhira2023}.  (b)~Field derivative $dM/dH_x$ versus $H_x$ of the data in (a) as indicated.  The field at the maximum of $dM/dH_x$ at each $T$ given by the vertical arrows is defined as the critical field $H_{\rm c1}$ at that $T$\@.}
\label{Fig_dMdHx(T)_[100]}
\end{figure}

\begin{figure}
\includegraphics [width=2.25in]{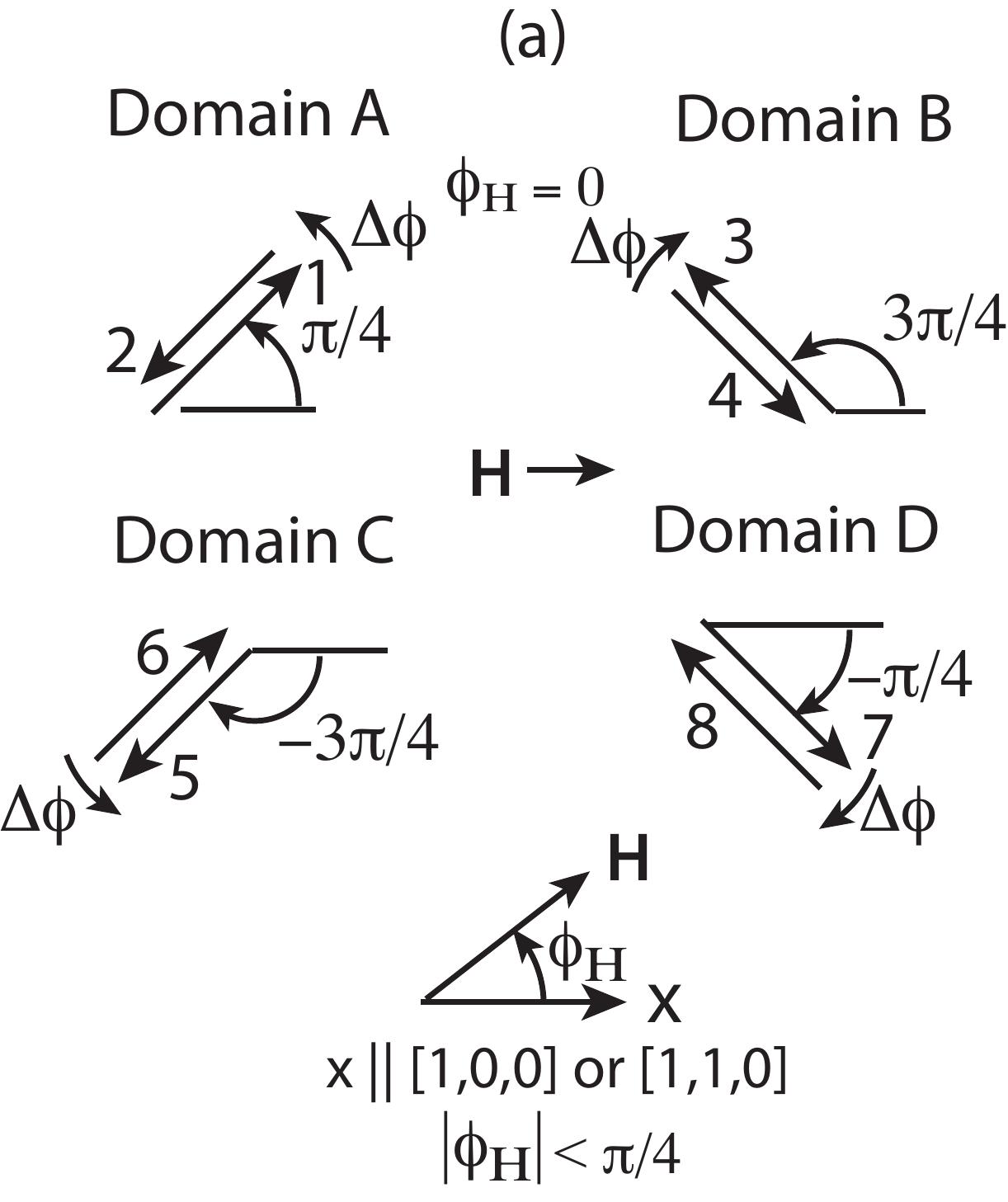}
\includegraphics [width=1.5in]{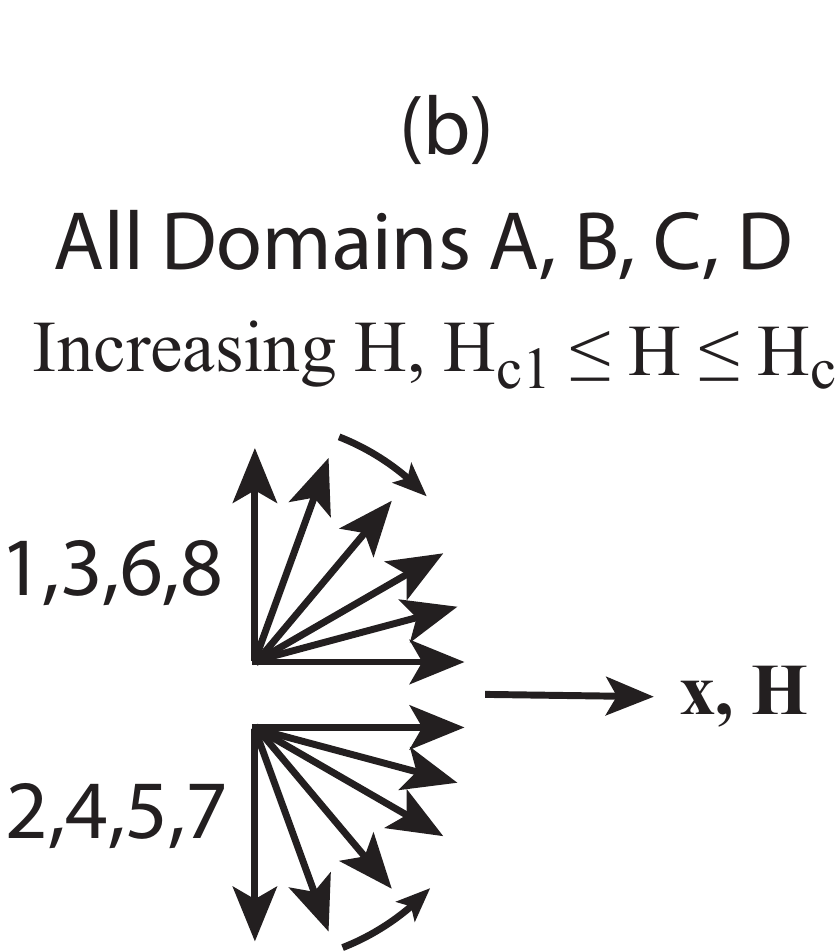}
\caption {(a)~Schematic diagram of the nearly-locked moment orientations in the $ab$-plane of adjacent antiparallel layers of moments along the $c$~axis in the four collinear A-type AFM domains A, B, C, and D and their magnetic field evolution with increasing $x$-axis field $H$ at low fields $H \leq H_{\rm c1}$ shown by arrows.  (b)~For $H_x>H_{\rm c1}$, each moment increasingly cants towards the increasing field as shown, until at the critical field $H_{\rm c}$ all moments are aligned with the field with \mbox{$\mu_x =\mu_{\rm sat} = gS\mu_{\rm B} = 7\,\mu_{\rm B}$.  From Ref.~\cite{Pakhira2023}.}}
\label{Figure_domain}
\end{figure}


The body-centered-tetragonal compound EuGa$_4$ contains Eu spin $S=7/2$ magnetic moments on the corners and body centers of the lattice.  It exhibits collinear \mbox{A-type} antiferromagnetic (AFM) order below $T_{\rm N} \approx 16$~K, where the Eu atoms are ferromagnetically aligned in the $ab$-planes and adjacent ferromagnetic (FM) planes along the $c$~axis are aligned antiferromagnetically~\cite{Nakamura2014, Nakamura2015, Nakamura2013, Zhu2022, Kawasaki2016, Yogi2013}.  It was also found that the magnetization $M$ in magnetic fields $H$ in the $ab$ planes exhibited positive curvature up to $H_{ab}\approx 6$~kOe, attributed to field-induced magnetic-moment reorientations, which decreased to zero at $T_{\rm N}$~\cite{Nakamura2013}. 

We subsequently studied the magnetic-field evolution of the AFM ground-state spin texture for $H\parallel [1,0,0]$ at $T<T_{\rm N}$  in detail, emphasizing the low-field region~\cite{Pakhira2023}, as shown in Fig.~\ref{Fig_dMdHx(T)_[100]}(a).  We found that although the \mbox{$c$-axis} magnetization $M_c$ increases linearly with the applied field $H$ as expected for an A-type antiferromagnet with the moments aligned in the $ab$~plane, an $ab$-plane $M_{ab}(H)$ behavior with positive curvature was observed at low $ab$-plane fields following by proportional behavior.  For $H\parallel [1,0,0]$ and $T=2$~K, this nonlinearity occurred from $H=0$ up to a critical field $H_{\rm c1} = 4.8$~kOe indicated in Fig.~\ref{Fig_dMdHx(T)_[100]}(b), above which $M_{ab}(H)$ attained a proportional behavior with slope $\chi_{ab}=\chi(T_{\rm N})$ as predicted by molecular-field theory (MFT)~\cite{Johnston2012, Johnston2015}.  The nonlinearity was found to vary significantly between the in-plane [1,0,0] and [1,1,0] field directions~\cite{Pakhira2023} as previously observed in Ref.~\cite{Nakamura2013}.

On the basis of the tetragonal structure of \eg\ and the $M_{ab}(H)$ isotherm measurements at $T=2~{\rm K}\ll T_{\rm N}$, we suggested a model for $T=0$~K~\cite{Pakhira2023} in which the A-type AFM structure consists of four-fold tetragonal domains orthogonal to each other in the plane in zero field that are equally populated by Eu spins as illustrated in Fig.~\ref{Figure_domain}(a).  Each domain contains pairs of ferromagnetic $ab$~planes that are aligned antiferromagnetically along the $c$~axis (A-type antiferromagnet).  We also assumed that within each physical AFM domain, the applied field $H_x$ can rotate the spins in the domain but not cause domain-wall motion.   In a body-centered-tetragonal lattice the magnetic-dipole interaction strongly favors moment alignment in the $ab$ plane rather than along the \mbox{$c$~axis}~\cite{Johnston2016}.  On application of a small magnetic field $H_x$, the moments in the domains were deduced from energy minimization to initially rotate to become nearly perpendicular to $H_x$ at a field $H_{\rm c1}$ where a maximum is observed in $dM/dH$ versus~$H$ as shown in Fig.~\ref{Fig_dMdHx(T)_[100]}(b), but where there is still a small tilting toward the field to yield the observed magnetization~$M_x$ at $H_{\rm c1}$.  For, $H_x > H_{c1}$, the system acts like a single domain and the moments start to cant towards the applied field direction as shown in Fig.~\ref{Figure_domain}(b), yielding a linear $M(H)$ behavior~\cite{Johnston2015}.  The magnetization saturates when all the moments become parallel to the applied field $H_x$ at a critical field $H_{ab}^{\rm c}$.

We observed similar behavior for the trigonal Eu-based compounds \emb, \ems, and \esa\ with A-type AFM order~\cite{Pakhira2020, Pakhira2021, Pakhira2021a, Pakhira2022}. For those cases we suggested a similar model for $T=0$~K to understand the $M_{ab}(H)$ behavior at $T\ll T_{\rm N}$ due to $ab$-plane field-induced Eu-moment reorientation in three trigonal domains equally populated by Eu spins~\cite{Pakhira2022b}. 

Here we extend the above zero-temperature theory to finite temperatures in order to model the $T$ dependence of the data for \eg\ in Fig.~\ref{Fig_dMdHx(T)_[100]} where $H_{\rm c1}$ depends strongly on $T$\@.   We find that the theoretical $M_{ab}(H)$ isotherms for temperatures in this range are in good agreement with the experimental data apart from the breadth of the transition $H_{\rm c1}(T)$, the source of which is not currently understood. 


In the A-type AFM state of \eg\ with tetragonal crystal symmetry and ferromagnetic (FM) layers of Eu spins aligned in the $ab$~plane, we infer  that fourfold FM domains occur in the $ab$~plane as shown in Fig.~\ref{Figure_domain}(a), given by
\bea
\phi_{\rm A} &=& \frac{\pi}{4} + \Delta\phi, \quad (0\leq \Delta\phi \leq \pi/4)\label{Eqs:phiABCD}\\
\phi_{\rm B} &=& \frac{3\pi}{4} - \Delta\phi,  \nonumber\\
\phi_{\rm C} &=& -\frac{3\pi}{4} + \Delta\phi, \nonumber\\
\phi_{\rm D} &=& -\frac{\pi}{4} - \Delta\phi, \nonumber
\eea
where the change $\Delta\phi$ depends on $T$ and the $ab$-plane magnetic field~$H_x$. 

The $ab$-plane anisotropy energies of the domains are given by~\cite{Buschow2003}
\bea
E_{\rm anis} = K_4 \cos[4\phi_n] \quad (n = {\rm A,~B,~C,~D}),
\label{Eq:Eanis}
\eea
where $K_4$ is the positive fourfold anisotropy constant and $\phi_i$ is the angle of the FM moments in a given $ab$-plane domain with respect to the positive $x$~axis, which is the direction of the applied field $H_x$.  Averaging over the angles $\phi_n$ in domains A--D gives
\bea
E_{\rm anis\,ave}(T) &=& - K_4\cos[4\Delta\phi],
\label{Eq:EanisAve}
\eea
where $\Delta\phi$ depends on both $T$ and $H_x$ as derived below. The value of $H_{\rm c1}$ is determined by $K_4$ according to \mbox{$H_{\rm c1} = \sqrt{8K_4/\chi_\perp}$}~\cite{Pakhira2023}.

According to molecular-field theory (MFT)~\cite{Johnston2015}, the magnetic susceptibility $\chi_\perp=\chi(T_{\rm N})$ for $H$ perpendicular to the moments in a ferromagnetically-aligned $ab$-plane domain  is independent of $T$ for $T\leq T_{\rm N}$, whereas  the magnetic susceptibility $\chi_\parallel$ parallel to the moments is zero for $T=0$~K and smaller than $\chi_\perp$ for all $T<T_{\rm N}$.  Therefore $\chi_\parallel$ can be ignored when minimizing the magnetic (free) energy for $T < T_{\rm N}$.

The magnetic energy of a moment in domain~$n$ in magnetic field $H_x$ is~\cite{Pakhira2023}
\bea
E_{{\rm mag\,}n} = -\mu_x H_x = -\chi_\perp H_x^2\sin^2(\phi_n).
\eea
In the regime \mbox{$0\leq H_x \leq H_{\rm c1}$}, the average over the four domains in Eqs.~(\ref{Eqs:phiABCD}) is
\bea
E_{\rm mag\,ave} = -\frac{\chi_\perp H_x^2}{2}[1+\sin(2\Delta\phi)],
\label{Eq:EmagAve}
\eea
where $0\leq\Delta\phi\leq\pi/4$.  This also gives
\bea
M_{x\,{\rm ave}} =-\frac{E_{\rm mag\,ave}}{H_x} =  \frac{\chi_\perp H_x}{2}[1+\sin(2\Delta\phi)],
\label{Eq:MxAve}
\eea
where $\chi_\perp=0.48$~cm$^3$/mol for \eg~\cite{Pakhira2023}.
Normalizing the energy by $K_4$, the total average energy including the magnetic contribution in Eq.~(\ref{Eq:EmagAve}) and the anisotropy contribution in Eq.~(\ref{Eq:EanisAve}) is 
\bea
\frac{E_{\rm ave}}{K_4} &=& \frac{E_{\rm anis\,ave}}{K_4} + \frac{E_{\rm mag\,ave}}{K_4}\nonumber\\
&=& - \cos(4\Delta\phi)\label{Eq:Etotal}\\
&& -\frac{\chi_\perp H_x^2}{2K_4}\left[1+\sin(2\Delta\phi)\right].\nonumber
\eea

For simplicity, we define
\bea
h_x = \frac{\chi_\perp H_x^2}{K_4},
\label{Eq:hx}
\eea
and Eq.~(\ref{Eq:Etotal}) becomes
\bea
\frac{E_{\rm ave}}{K_4} &=& - \cos(4\Delta\phi) -\frac{h_x}{2}\left[1+\sin(2\Delta\phi)\right].
\label{Eq:Etotal2}
\eea

\begin{figure}
\includegraphics [width=3in]{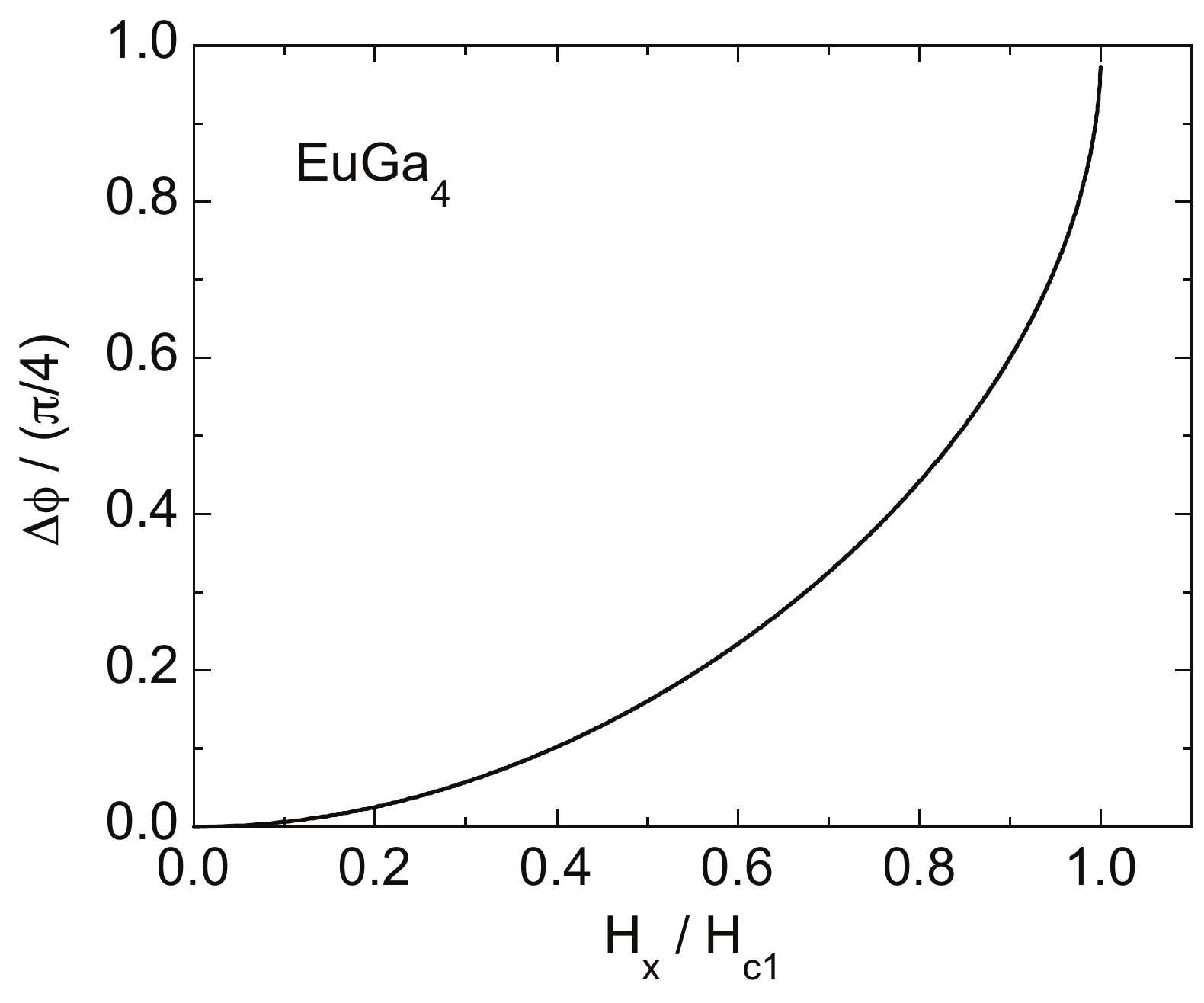}
\caption {Universal variation for $T\leq T_{\rm N}$ of the reduced change in angle $\Delta\phi/(\pi/4)$ vs reduced field $H_x/H_{\rm c1}$ obtained using Eq.~(\ref{Eq:DeltaPhi4a}).  }
\label{Fig_DphiVsHx}
\end{figure}

Minimizing $E_{\rm ave}/K_4$ with respect to $\Delta\phi$ gives~\cite{Pakhira2023}
\bea
\Delta\phi = \frac{1}{2}\arctan\Bigg(\frac{h_x}{\sqrt{64-h_x^2}}\Bigg).
\label{Eq:DeltaPhi2}
\eea
Thus $\Delta \phi=\pi/4$ when $h_x$ attains the value 8, which is denoted as $h_{\rm c1}$.  This is the value at which all moments become nearly perpendicular to $H_x$ according to Fig.~\ref{Figure_domain}(a), which is   defined above as $H_{\rm c1}$, apart from a slight canting towards the field to give the observed small magnetization at $H_{\rm c1}$.  For larger fields the moments increasingly cant towards $H_x$ as shown in Fig.~\ref{Figure_domain}(b) until saturation is reached at the critical field $H_{ab}^{\rm c}$.  Therefore we write Eq.~(\ref{Eq:DeltaPhi2}) as 
\bse
\bea
\Delta\phi &=& \frac{1}{2}\arctan\Bigg(\frac{h_x}{\sqrt{h_{c1}^2-h_x^2}}\Bigg)\label{Eq:DeltaPhi3}\\
&=&\frac{1}{2}\arctan\Bigg[  \frac{h_x/h_{\rm c1}}{\sqrt{1-(h_x/h_{\rm c1})^2}}   \Bigg].\label{Eq:DeltaPhi3A}
\eea
\ese

Using Eqs.~(\ref{Eq:hx}) and~(\ref{Eq:DeltaPhi3A}), we obtain
\bse
\bea
\Delta\phi &=& \frac{1}{2}\arctan\left[\frac{(H_x/H_{\rm c1})^2}{\sqrt{1-(H_x/H_{\rm c1})^4}}\right](H_x\leq H_{\rm c1}),\nonumber\\
\label{Eq:DeltaPhi4a}\\
M_x&=&\chi_\perp H_x \quad (H_{\rm c1}\leq H_x \leq H^{\rm c}_{c}),\label{Eq:DeltaPhi4b}\\
M_x &=& M_{\rm sat} = N_{\rm A}gS\mu_{\rm B}\quad (H_x\geq H^{\rm c}_{c}),
\eea
\ese
where $N_{\rm A}$ is Avogadro's number, $g=2$, $S=7/2$, and $H^{\rm c}_{c}$ is the $c$-axis critical field which is 72~kOe at $T=2$~K and 38~kOe at $T=14$~K~\cite{{Nakamura2013}}.

\begin{figure}[ht]
\includegraphics [width=2.8in]{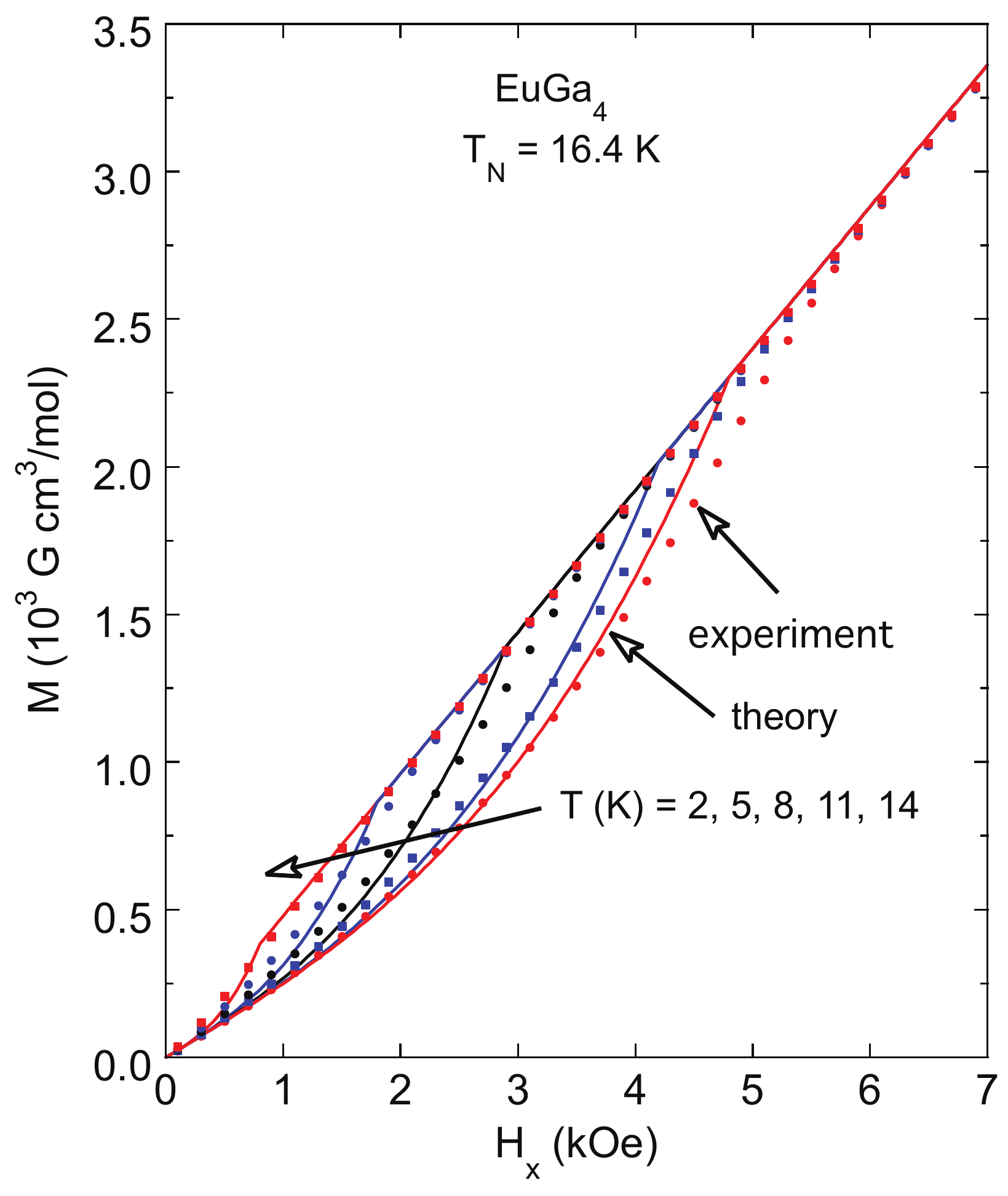}
\caption {Magnetization~$M$ versus applied field $H_x$ for $H\parallel [1,0,0]$ at the temperatures indicated.  The data symbols are the measured data at each temperature as indicated, and the solid curves are the theoretical fits to the respective $M(H)$ isotherm data.  The fits to the data are quit good except near the critical fields $H_{\rm c1}$ at which the $M(H_x,T)$ data at each temperature become proportional to $H_x$.  The high-field slopes of the data at each temperature have the same value $\chi=\chi(T_{\rm N})$, in agreement with the prediction of molecular-field theory. }
\label{Fig_M_vs_H_Hc1_Thy_Expt}
\end{figure}

Figure~\ref{Fig_DphiVsHx} shows $\Delta\phi$ normalized by $\pi/4$ versus the ratio $H_x/H_{\rm c1}$ from 0 to 1 calculated using Eq.~(\ref{Eq:DeltaPhi4a}), which exhibits a smooth increase in $\Delta\phi$  from 0 to $\pi/4$ over this range.

The experimental $M_{ab}(H,T)$ data for \mbox{$H_x \parallel [1,0,0]$} from Fig.~\ref{Fig_dMdHx(T)_[100]}(a), along with the theoretical $M_{ab}(H_x)$ data calculated using the $H_{\rm c1}(T)$ data in Fig.~\ref{Fig_dMdHx(T)_[100]}(b) and Eqs.~(\ref{Eq:MxAve}), (\ref{Eq:DeltaPhi4a}), and (\ref{Eq:DeltaPhi4b}) (solid curves), are shown in Fig.~\ref{Fig_M_vs_H_Hc1_Thy_Expt}.  The theory quantitatively reproduces the experimental $M(H)$ data at low and high fields, but does not reproduce the curvature in the experimental data near $H_{\rm c1}$.  The reason for the latter behavior is not clear at present.  A qualitatively similar but quantitatively larger discrepancy between the theoretical and experimental data near $H_{\rm c1}$ taken at $T = 1.8$~K was observed earlier for \emb\ and \ems, where the measurement temperatures were  $T \sim 0.27~T_{\rm N}$ and $T \sim 0.23~T_{\rm N}$, respectively.  The discrepancy is smaller for \eg\ because the minimum measurement temperature here was $T = 2~{\rm K} \approx 0.13\,T_{\rm N}$. 

In summary, an anomalous positive curvature at small fields is observed in the $M_{ab}(H,T)$ isotherms at $T< T_{\rm N}$ for the tetragonal square-lattice antiferromagnet \eg, which exhibits A-type AFM order below $T_{\rm N} = 16.4$~K with the moments aligned in the $ab$~plane.  A theory was presented that fits these $T$-dependent $M_{ab}(H)$ isotherm data rather well based on the occurrence of fourfold AFM domains.  The same theory could also be used to fit  $T$-dependent $M_{ab}(H)$ isotherms below $T_{\rm N}$ for trigonal \mbox{A-type} antiferromagnets such as \emb\ and \ems.  An interesting avenue for future research would be to determine the source of the transition widths at $H_{\rm c1}(T)$ as evident in Fig.~\ref{Fig_dMdHx(T)_[100]}(b).

\vspace{-0.2in}

\acknowledgments

This research was supported by the U.S.~Department of Energy, Office of Basic Energy Sciences, Division of Materials Sciences and Engineering.  Ames National Laboratory is operated for the U.S. Department of Energy by Iowa State University under Contract No.~DE-AC02-07CH11358.

\end{document}